# Aqueous metal-organic solutions for YSZ thin film inkjet deposition

C. Gadea*[a], Q. Hanniet[a], A. Lesch[b], D. Marani[a], S. H. Jensen[a], V. Esposito[a]

a: DTU Energy, Technical University of Denmark, Risø Campus, Frederiksborgvej 399,

DK - 4000 Roskilde, Denmark

b: Laboratoire d'Electrochimie Physique et Analytique, EPFL Valais Wallis, Rue de l'Industrie 17,

CH-1950 Sion, Switzerland

## Abstract

Inkjet printing of 8% $Y_2O_3$-stabilized $ZrO_2$ (YSZ) thin films is achieved by designing a novel water-based reactive ink for Drop-on-Demand (DoD) inkjet printing. The ink formulation is based on a novel chemical strategy that consists of a combination of metal oxide precursors (zirconium alkoxide and yttrium salt), water and a nucleophilic agent, *i.e.* n-methyldiethanolamine (MDEA). This chemistry leads to metal-organic complexes with long term ink stability and high precision printability. Ink rheology and chemical reactivity are analyzed and controlled in terms of metal-organic interactions in the solutions. Thin dense nanocrystalline YSZ film below 150 nm are obtained by low temperature calcination treatments (400-500 °C), making the deposition suitable for a large variety of substrates, including silicon, glass and metals. Thin films and printed patterns achieve full densification with no lateral shrinkage and high ionic conductivity.

# 1. Introduction

Solid oxide fuel cell (SOFC) is a mature technology where materials' processing is the key to improve performances and to boost commercially viability. In the past decades, research has been focused on lowering the operating temperatures to make SOFC more durable and financially competitive [1–5]. One promising approach is the reduction of the electrolyte layer thickness [6] to lower ohmic losses [7]. Several processes have been considered for the fabrication of thin-film electrolytes for SOFC. While the deposition of films < 10 µm using spray coating [8], tape casting [9], electrophoretic deposition (EPD) [10] and screen printing [11] is technically challenging, more advanced techniques, such as pulsed laser deposition (PLD) [12], chemical vapor deposition (CVD) [13] or spray pyrolysis [14] can produce thin films with fine microstructure and compositions. However, the more advanced techniques are generally difficult to scale up and are not cost effective. Additive manufacturing such as Drop-on-Demand (DoD) inkjet printing represents a relatively new deposition technology with unique advantages including customizable patterns and shapes. This is of growing interest for SOFC since it is ideal for controlled thin films deposition and shapes in multicomponent layer systems such as SOFCs [15–18]. The implementation of printheads with several hundred individually addressable nozzles, ejecting highly reproducibly picoliter droplets ($10^{-12}$ liter) of well-defined ink compositions, enables a comfortable route for up-scaling from the prototype development to the industrial fabrication level.

In a previous study, we deposited a 1.2 µm thick YSZ electrolyte layer by inkjet printing a colloidal based solution (*i.e.* a stabilized oxide particles suspension in an aqueous medium) on a 9×9 cm² porous nickel oxide/yttria-stabilized zirconia (NiO/YSZ) anode [15]. The YSZ printing showed a high potential in terms of both processing flexibility and electrochemical performance of the cell [15]. However, inkjet printing of colloids shows some limitations and it does not suit thin film deposition << 1 µm. Indeed, the minimum layer thickness depends on the particle size in the colloid as the total thickness is the result of the particles' packing and the number of layers needed to achieve a full coverage of the substrate [15]. The drying behavior of such ink is also a critical to avoid "coffee ring effects" leading to non-continuous films [19]. Moreover, stabilizing nanoparticles below 100 nm to achieve thinner films is challenging due to the large amount of stabilizers and additives needed to cover the particles´ high surface area [20]. Finally, the consolidation of ceramic-powder based layers occurs at high temperatures (e.g. 1300°C for ca. 100-500 nm YSZ powders [21]). This step imposes limitations on the substrates selection to be used, excluding silicon, metals, and nanostructured substrates, that are usually selected for thin-film based SOFCs made by PVD and CVD techniques [14]. Alternatively to colloidal suspensions, particle-free, chemically reactive solutions contain only dissolved precursor salts or molecules. After printing, the precursors are converted into solid particles based on thermal, chemical or photochemical post-processing steps. Reactive inks for the deposition of metal oxide films usually contain dissolved precursors, such as metal alkoxide molecules in alcoholic solution [22–24]. Reactive inks can be found in various forms [25–28]. The sol-gel process consists in a combination of reactions (hydrolysis and condensation) of the metal alkoxide which leads to the formation of homogeneous oxo/hydroxo-polymeric matrix at room temperature [29]. Due to such a structure, sol gels do not need mass diffusion steps such as sintering at high temperatures to become dense; the precursor only requires a calcination step to remove by-products and additives and to crystallize the phase from the polymeric form. Alkoxide precursors are generally highly reactive, *e.g.* towards water, leading to an unavoidable premature gel formation. Therefore, the full control of sol gel precursor reactivity is crucial, especially for inkjet, where inorganic polymerization can lead to irreversible clogging of nozzle or reservoir. Inkjet printing of sol-gel based $TiO_2$ films was stabilized by adding a nucleophilic ligand (or chelating agent), N-methyldiethanolamine (MDEA) [30]. MDEA stabilizes the alkoxide precursors by coordinating the metallic cation from water and thus inhibiting the hydrolysis/condensation reactions [31]. While several studies describing the deposition of sol-gel based YSZ electrolytes for SOFC are reported in literature using other processing methodologies such as spin coating [32,33] or dip coating [34], to the best of our knowledge, no research on inkjet printing of YSZ sol-gel based ink with nanometric features has been reported so far. The use of reactive inks for SOFC electrolytes was solely done by Wang *et al.* [21] for the deposition of Gadolinium-doped Ceria (CGO) thick layers (10 µm and thicker) on porous NiO/YSZ and YSZ/CGO substrates and no analysis of inks rheology and stability of YSZ reactive inks are available in the literature. In this work, we fill this gap for environmental-friendly water-based reactive YSZ ink, demonstrating its long term stability, suitable printability and versatility in depositing on several substrates by using both thermal and piezoelectric DoD inkjet printers [35].

## 2. Experimental

### 2.1. Ink preparation

The YSZ ink formulae were elaborated to produce 8 mol% YSZ (i.e. $0.08Y_2O_3$-$0.92ZrO_2$) starting from zirconium (IV) propoxide ($Zr(OPr)_4$, 70 wt.% in propanol, Sigma Aldrich) as zirconium precursor and yttrium (III) nitrate hexahydrate ($Y(NO_3)_3,6H_2O$, 99.8% trace metals basis, Sigma Aldrich) as dopant. N-methyldiethanolamine (MDEA, Sigma-Aldrich) was used as the nucleophilic agent. Ink preparation was carried out under argon atmosphere to avoid any undesired reactions between zirconium alkoxide and ambient water. MDEA and $Zr(OPr)_4$ were mixed to chelate alkoxide molecules, followed by the addition of ethanol. In a separate reactor, yttrium nitrate salt was dissolved in water and kept under stirring until complete salt dissolution. The aqueous yttrium salt solution was then slowly and dropwise added to the solution containing the chelated propoxide and mixed for 5 min. To screen the ink properties and identify an optimized formulation for printing, several parameters were varied: the water/ethanol volume ratio from 100/0 to 0/100, molar ratio $r$ = [MDEA]/ [Zr(OPr)$_4$] and the concentration $c$ of zirconium from 0.1 to 0.2 g/ml. The hydrolysis ratio $h$ = [H$_2$O]/ [Zr(OPr)$_4$] was kept constant for all inks ($h$ = 8). Herein, inks are labeled as **Ink *r-c-X***, where $r$ refers to the molar ratio of complexation, $c$ to the zirconium concentration, and $X$ to the water volume percentage.

*2.2. Ink characterization*

Rheological properties of the inks were explored using an Anton Paar Rheometer (MCR 302), in rotational mode and at a constant temperature of 21 °C. A plate-plate measuring system was used with a diameter of 50 mm (PP50) and at gap distance of 0.5 mm. A solvent trap was used to avoid ethanol evaporation. The experiments were performed using three steps of pre-treatment: the first one at a shear rate of 1 s$^{-1}$ for 1 min followed by 1 min at rest (shear rate = 0 s$^{-1}$), and the third one at 1 s$^{-1}$ for 1 min. The pre-treatment steps were adopted to remove any effects due to the sampling and loading of the inks. Flow curve measurements were conducted in step mode using 60 steps with a waiting time of 10 s. This procedure allows the samples to reach equilibrium and avoid possible transient effects. The shear rates investigated range from 1 s$^{-1}$ up to 100 s$^{-1}$. The viscosity was measured over time, from a few hours after preparation up to 360 h (15 days) in order to estimate the long term ink stability. Ink surface tension was measured using a bubble pressure tensiometer (BP 50, Krüss), and density was evaluated by weighing 10 ml of ink.

The printability of the inks was determined by calculation of their $Z$ numbers defined as:

$$Z = \frac{\sqrt{\sigma \rho a}}{\eta} \qquad (1),$$

where $\eta$ is the ink viscosity (mPa·s), $\sigma$ is the surface tension (mN/m), $\rho$ is the density (g/cm$^3$), and $a$ is the characteristic length (µm) [36]. The parameter $a$ is typically taken as the nozzle diameter [36]. The specifications of the printers are given in Section *2.3*.

*2.3. Thin films deposition and characterization*

Substrates used in this work were Si (100), quartz (Crystal GmbH), polycrystalline alumina (Keral96, Kerafol), and sintered nickel oxide/yttria stabilized zirconia (NiO/YSZ) made in-house. For the latter, the layers were produced by tape-casting, co-laminated as green materials at *ca.* 150°C [37], and sintered at 1300°C for 6 h in air. The substrates had a 10-15 µm thick anode of NiO/YSZ cermet laminated to a ~300 µm thick NiO/YSZ support layer. The ratio between Ni and YSZ was 40/60 vol. % both for the support layer and the active electrode layer. $ZrO_2$ stabilized with 8 mol. % $Y_2O_3$ was used for the anode layer while $ZrO_2$ stabilized with 3 mol. % $Y_2O_3$ was used for the support layer.

A commercially available thermal printer (HP Deskjet 1010) was used as a representative thermal inkjet printing unit. The printer was modified to allow printing on flat and thick substrates (rigid or flexible) [15]. A compatible cartridge (HP 301 black) providing a 600 × 300 dpi resolution was cleaned to remove the original black ink and used to print the sol-gel inks. The cartridge contains 336 nozzles each having a diameter of 20 µm. A high resolution piezoelectric based X-Serie CeraPrinter (Ceradrop) was used as a representative piezoelectric printing unit. Both working principles and main differences in terms of required ink properties are listed elsewhere [30,35]. This printer is equipped with a fiducial camera to observe droplet ejection and calculate droplet velocity. Disposable DMP cartridges DMC-11610 (Fujifilm Dimatix) containing 16 individually addressable nozzles (nozzle diameter = 21.5 µm) were used with a custom-made jetting waveform. In brief, the waveform consisted of a baseline at 5 V, a 5 µs segment at -10 V to fill the nozzle with ink from the reservoir and a droplet ejection segment at 40 V for 5 µs with well-defined slopes between each segment. The jetting frequency was varied between 1 and 10 kHz. The obtained droplet velocity was 5 m/s. Each print made with both printers consisted of a single layer.

The evolution of the crystallographic phase for each material was investigated by x-ray diffraction (XRD) using solid state detector, 0.01° step size, and 0.5 s step time (Bruker D8, Cu Kα radiation). Diffraction patterns of powder resulting from the ink calcination at various temperatures were recorded. The crystallite size was determined using Scherrer equation [38] with internal reference. The value of the Cu-Kα radiation used for determination is 0.15406 nm and the *k* value is equal to 1.

Thermal analysis (DTA/TG) was performed on inks using a STA 409 PG (Netzsch) at a constant rate of 10 K min$^{-1}$ under air between 25 and 700°C.

As result of XRD and TG-DTA characterizations, the calcination of the printed patterns on different substrates was carried out at 500 °C for 1 h in air using a chamber furnace, with a heating ramp of 100 °C/h after 2 intermediate steps at 90 °C and 120 °C of 1 h each. Morphological characterization of the printed films was investigated by using scanning electron microscopy (SEM, SUPRA 35, Zeiss). Film surface and thickness characterizations were carried out by using a high resolution 3D profilometer (Cyberscan, Vantage, using a LT9010 detector, a non-contact laser based method to analyze surfaces). The measured film dimension was 5×5 mm$^2$ and the measurement step size was 50 µm for both *x*- and *y*-axis.

Electrochemical impedance spectroscopy (EIS) was performed in a symmetric 2-electrode configuration by using a Solartron 1260 frequency response analyzer over the frequency range of 1 Hz-86 kHz. As also reported in previous papers, such a configuration can be used to characterize the in-plane ionic conduction at the thin film samples on dielectric substrates [39,40]. Due to the high impedance of the thin films in the lateral direction, 200 mV AC voltages were used for the acquisition. Symmetric silver electrodes were painted on the surface to measure the lateral conductivity of the calcined samples. The electrodes dimensions were measured to be ca. 1 mm distance and 2 mm in width. EIS measurements were performed in the temperature range of 525-750 °C in synthetic air flow (10 sccm).

## 3. Results and Discussion

*3.1. Ink properties*

The ability of an ink to be jetted is controlled by several intrinsic properties such as density, viscosity and surface tension of the fluid [41–44]. The printability number $Z$ (Eq. 1), based on these physical properties, has been proposed and printability intervals are well-defined in the literature. For instance, Derby has suggested an interval between 1 and 10 based on simulations [36], whereas Jang *et al.* [45] defined an interval between 4 and 14, using experimental observations on jetting of solvent mixtures. In our previous study, we showed that $TiO_2$ reactive sol-gel inks can be reproducibly printed with a printability number close to 1, where stability and control towards hydrolysis/condensation reactions in a reactive ink is the crucial factor. [30] Adding MDEA as nucleophilic ligand to the solution a high stability of the ink for printing can be achieved. The ligand, however, is the most viscous component in the ink and reduces the printability number. As a result, the optimal amount of ligand to inhibit the hydrolysis/condensation reactions, while keeping a suitable printability number, has to be determined in relation to the metal-organic system and the media in use.

To identify the optimal MDEA content required for stability, the physical properties that define the printability number were investigated for a series of ink samples having different molar complexation ratios r = [MDEA]/ [Zr(OPr)4] from 9 to 15. Preliminary tests with r < 9 (results not shown), exhibited extremely quick gelation of the inks, and were not investigated further. Too high concentration of MDEA, *e.g.* r > 15, was not considered to avoid excessive shrinkage of the final printed pattern and polymerization of the ligand [30]. Moreover, inks having only water as solvent system were investigated, *i.e.* ink r-0.1-100, to understand the water-precursors interaction, avoiding any effects of the co-solvents on the ink stability. Figure 1 shows the flow curves for four of such water-based inks, 1 h after their preparation.

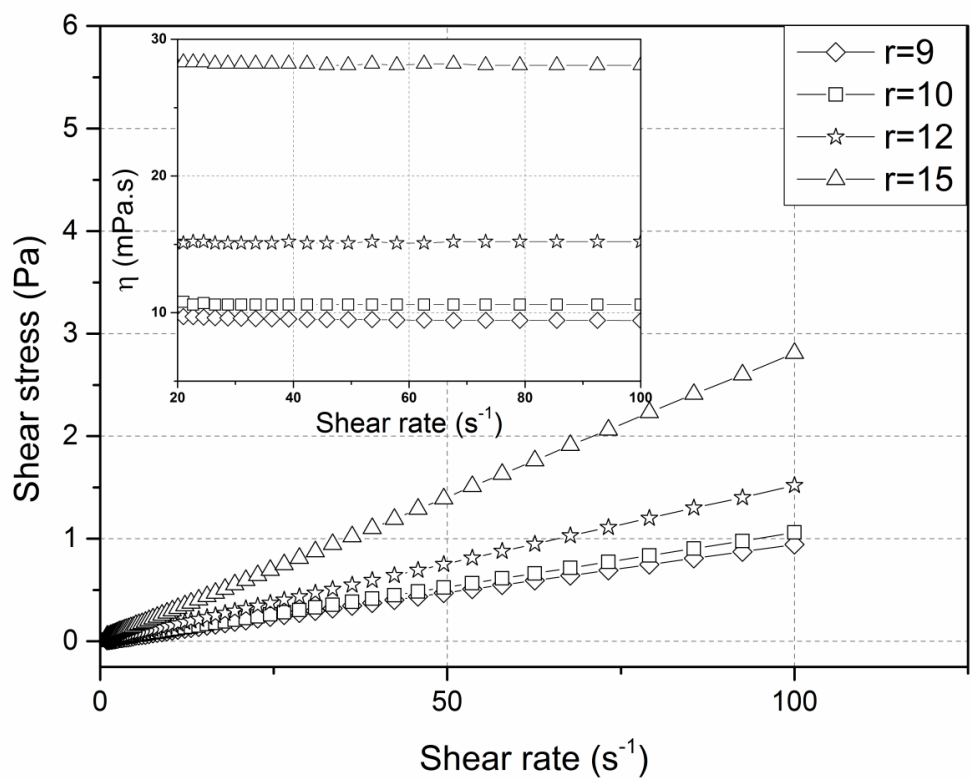

**Figure 1: Flow curve and viscosity curves (inset) of the four developed inks measured 1 h after preparation.**

The corresponding viscosity curves are presented in the inset. All prepared inks showed a Newtonian behavior, *i.e.* a constant viscosity value for the considered shear rate range [30]. Table 1 lists the viscosity values calculated from the slope of the flow curves in Figure 1, along with the measured surface tensions and printability. This analysis clearly indicates that viscosity values increased with the amount of MDEA in the ink as a possible effect of polycondensation of the alkoxide-ligand complexes [30,46]. However, increasing the amount of MDEA did not affect the surface tension of the inks, which fluctuated by a few mN/m. The calculated printability numbers of inks aged for 1h from Table 1 indicate that all inks except *ink **15-0.1-100*** exhibited a printability number within the $1 < Z < 10$ range [36].

The long-term stability of the inks at different concentrations was evaluated following the evolution of printability over time. Figure 2 shows stability tests carried out with the ink samples and reported as printability as a function of *r* at different ageing times, from 1 to 360 h.

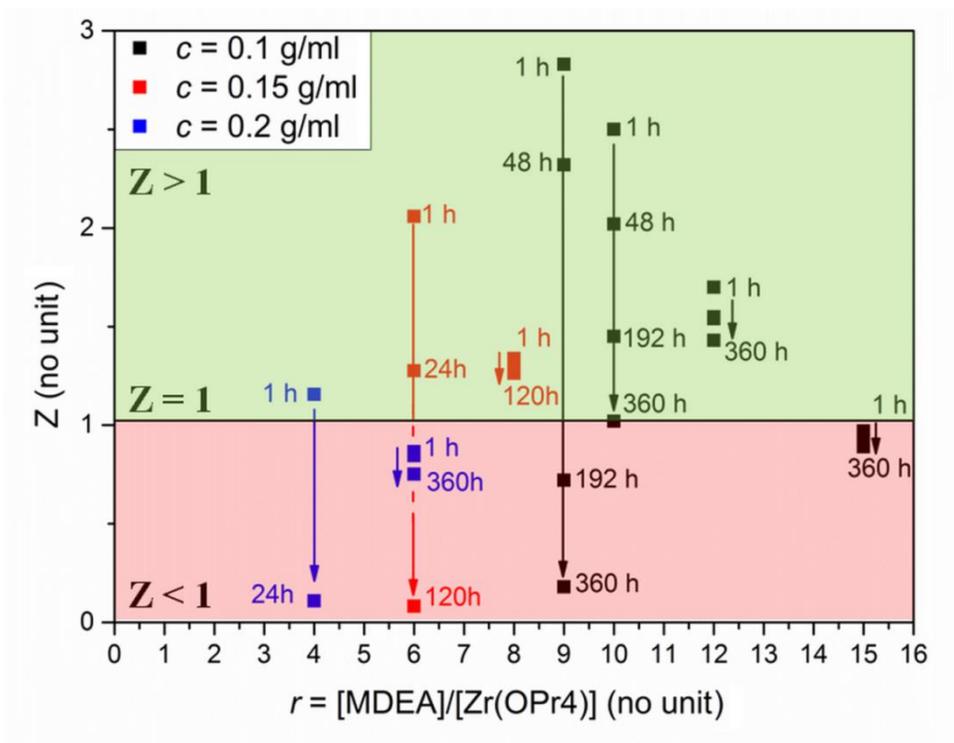

**Figure 2: Inks printability at different concentrations and aging time *versus* complexation ratio.**

A green area in the plot represents the printable zone, which corresponds in this study to the viscosity range of inks having a printability number equal or lesser than 1, while a red area represents the viscosity range where inks are theoretically not printable ($Z < 1$). In this long-term stability study, the decrease in $Z$ value is strongly linked to a viscosity increase caused by a less effective hydrolysis/polymerization inhibition over time. The frontier viscosity value between the 2 zones ($Z = 1$) is based on typically observed values for surface tension (31 mN/m) and density (1 g/cm$^3$) for this type of inks (see also Table 1). As the hydrolysis ratio $h$ was kept constant for all the inks, different complexation ratio intervals were studied for each concentration. For the inks with $c = 0.2$ g/ml (blue dots in figure 2), ***ink 4-0.2-100*** was in the printability range for ca. 1 h, before going out of it with a too low Z after 24 h of aging. Despite being all the time outside of the printability range, the ***ink 6-0.2-100*** gave a very stable $Z$ value. For inks with $c = 0.15$ g/ml (red dots in figure 2), ***ink 6-0.15-100*** was printable for approximately 24 h before its printability number dropped below 1. On the other hand, ***Ink 8-0.15-100*** exhibited a stable printability number > 1 over 120 h. For $c = 0.1$ g/ml (black dot in Figure 2), data showed that inks made with low MDEA content ($r = 9$ and 10) exhibited a rapid decrease of printability number after a few days of aging. Finally, at higher MDEA content ($r = 12$ and 15), ink viscosities showed a remarkable long-term stability but only ***ink 12-0.1-100*** remains printable after 360 h.

In summary, these observations indicate clearly that long term-stability was more difficult to achieve at high concentrations of zirconium in the solution. As the complexation ratio is reduced due to concentration increase, the alkoxide is less and less "protected" towards water, leading to YSZ polymerization which increases viscosity over time [30]. However, improved stability might be reached by increasing the complexation ratio $r$, and decreasing the hydrolysis ratio $h$. At low concentration ($c = 0.1$ g/ml), the same effect is observed for low complexation ratios ($r = 9$ and 10), which resulted in unstable inks. On the other hand, high MDEA content in ***ink 15-0.1-100*** had the undesirable effect to decrease the printability. As also observed and demonstrated for Ti-based inks, the increase of viscosity for high MDEA concentrations can be the consequence of alkoxide-ligand complex polycondensation [30,46]. As a result of this detailed analysis, we defined a proper long-term stabilization for the complexation ratio with an optimal $r$ of 12. As a consequence ***ink 12-0.1-100*** achieved a suitable chelate effect on the alkoxide and preserved the printability in the long term.

It is worth noticing from Table 1 that ink surface tension $\sigma$ was very slightly affected by MDEA content over time, making viscosity the main physical property controlling the printability. Another important feature of such inks is that the printability can be further refined and improved by diluting the solvent with a less viscous compound. We proposed ethanol, which not only reduces viscosity but also can improve the ink drying rate and reduce the surface tension. Such a use of alcohols in general is crucial in surface chemistry, especially when depositing water based inks. For example, the inks developed in this study contain propanol from the alkoxide, which reduces the high surface tension of water (72.8mN/m at 20°C) to much lower values (ca.31 mN/m, see Table 1). Such decrease improves the substrate wettability to maximize the substrate coverage, leading to homogenously printed continuous thin films [47]. The key point in tailoring the solvent composition is then not only in the ink jetting process, but also in depositing the ink on substrates with different surface free energies and in improving the drying rate.

As result of such an analysis, a series of optimized inks with $r = 12$ and $c = 0.1$ g/ml, with different water/ethanol ratios (from $X=0$ to 100) were synthesized to evaluate the most promising ink for inkjet printing. The measured physical properties of the inks at different aging times are reported in Table **2**. Such values indicate that all the inks were within the defined printable range in the considered aging time frame. As expected, both viscosity and surface tension were significantly decreased as the ethanol content increases, rising up the inks printability after 360 h of aging from 1.4 for $X = 100$ to 2.7 for $X = 0$. Moreover, the surface tension of the ink spans from 31 to 24 mN/m, when $X$ is reduced. Such decrease in surface tension results in better coverage of substrates, and opens the possibility of using a larger choice of substrates. However, while increasing the volume of ethanol can be beneficial for droplet formation and wettability the water/ethanol ratio can also have a strong influence on the drying, where too fast drying can lead to clogging at the nozzle. Particularly for the presented YSZ inks, clogging for $X < 40$ vol% inks (*i.e.* inks with more than 60 vol% ethanol) was observed. On the basis of these observations we conclude that from the presented matrix of inks, the ink with formula *ink 12-0.1-60* is the most suitable ink for printing.

*3.2. Ink jettability*

The jettability of *ink 12-0.1-60* was studied first by using the piezoelectric printer.

Figure 3 (a) shows a droplet of fresh *ink 12-0.1-60* after ejection at a firing frequency of 1 kHz at different distances from the nozzle, observed with the camera mounted on the printer.

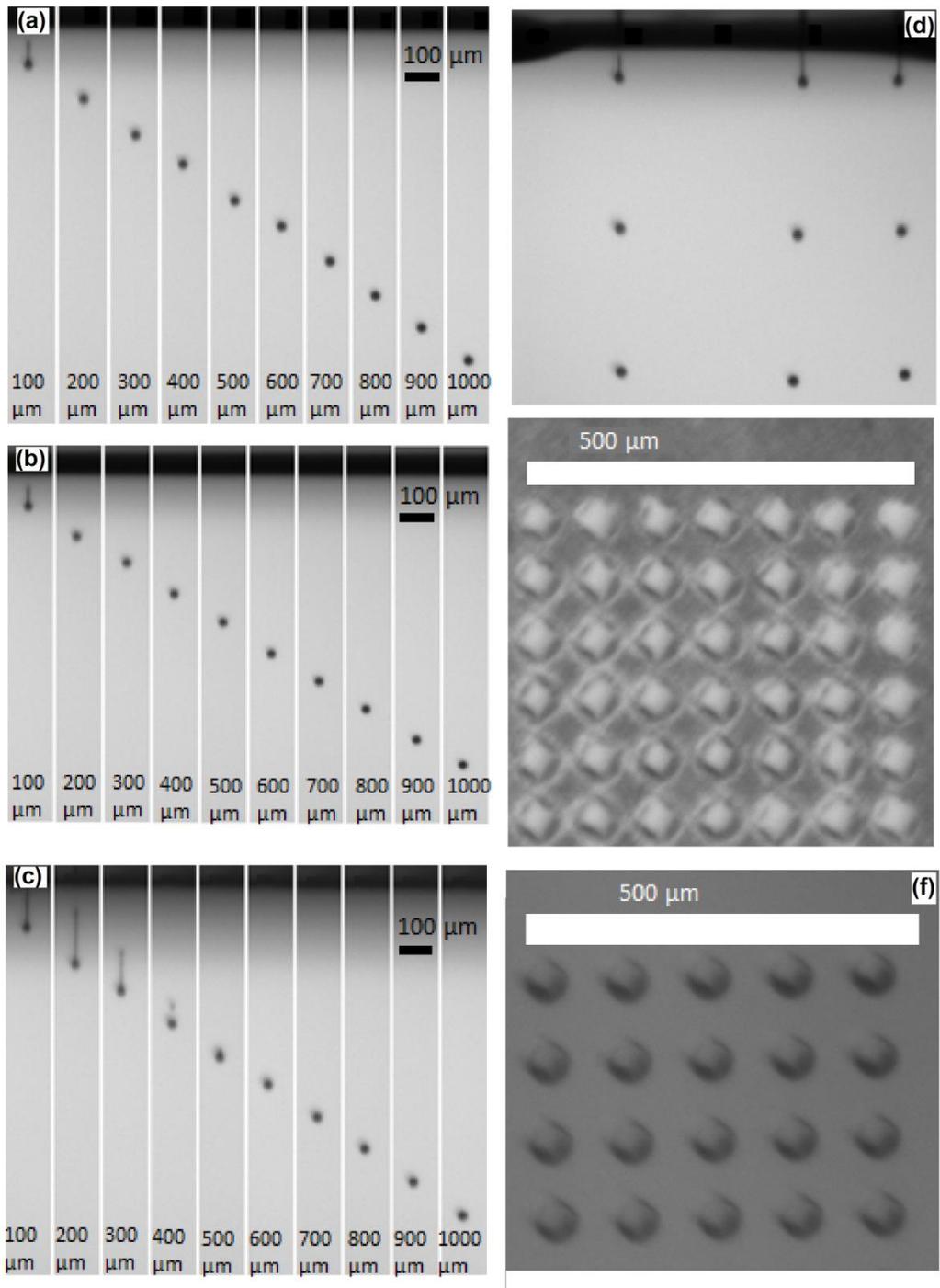

**Figure 3:** Droplet at different nozzle distances (frequency=1 kHz) of ink 12-0.1-60 as prepared (a), aged for 48h (b), and aged for 1200h (c), followed by droplets of fresh ink 12-0.1-60 (frequency = 10 kHz) (d) and optical pictures of an array of fresh ink 12-0.1-60 printed on NiO/YSZ (e) and alumina (f).

At a distance of 100 µm, a filament with a round head is formed, due to the ejection process and the surface tension difference between the ink and the nozzle plate. The droplet was not completely spherical at this stage. However, in the range 200-1000 µm, the ejected droplet exhibited a remarkable stable spherical shape in the considered nozzle-droplet distance range. Moreover, no satellite droplets could be observed in the whole distance range, indicating high quality printability. Such a result is in accordance with the ability to print when the printability (here, $Z = 1.7$) is in the range $1 < Z < 10$. Figure 3 (b) shows the result of the same analysis carried out with *ink 12-0.1-60* aged for 48 h on the shelf ($Z = 1.5$) with similar results than for the fresh ink, verifying the reliability of the 48 h-term stability and printability at the jetting. With the conditions of jetting used in the experiment, the fully formed droplets showed a stable diameter in the range of 30-32 µm up to 48 h. Finally, Figure 3 (c) shows the jettability of *ink 12-0.1-60* aged for 1200 h (50 days). This latter ink required a minor adjustment of the waveform, where just the nozzle filling voltage was slightly modified from -10 to -7 V while keeping the general waveform shape and the firing voltage the same. A long tail, which was not observed for fresher inks, could be noticed above the droplet in the distance range 100-300 µm, followed by a breaking of the tail at 400 µm, resulting in a main droplet and satellite droplets. However, satellites merged at 500 µm to form again a spherical and stable droplet in the range 500-1000 µm. Such a change of the jetting was attributed to a slight viscosity increase in the long term (18.1 mPa.s, $Z = 1.3$ at 1200 h). Despite the presence of a long tail and satellite droplet in the first hundreds of micrometer from the nozzle, this observation demonstrates the outstanding stability of the ink after a long term aging of 50 days. Yet, the jetted droplet diameter was comparable to the one measured for the fresh inks and similar droplet shape stability was also observed in the 48 h aged ink at increased firing frequency, from 1 kHz to 10 kHz (Figure 3(d)).

A series of droplets was deposited on two different substrates with different microstructures, *i.e.* dense alumina and porous NiO/YSZ anode, using a lateral droplet spacing (*i.e.*, distance between the center of two adjacent droplets) of 100 µm. Figure 3 (e)-(f) shows the resulting patterns after deposition. For both substrates, the deposited droplets appear as circular dots and exhibited a reproducible diameter of *ca.* 52±2 µm. This further illustrates the versatility to deposit this ink on very different surfaces at the micrometric scale.

*3.3. Phase and microstructure characterization*

The deposition of reactive ink based on sol-gel leads to the formation of an amorphous polymeric matrix. This material results from the hydrolysis and condensation reactions of the zirconium alkoxide, the yttrium based dopant, as well as solvents and agents such as MDEA. Therefore, a calcination step is necessary to remove the organics and crystallize the deposited layer. Previous thermal analysis (TG/DTA) performed on the sol-gel system using MDEA [30] showed that the mass loss relative to solvent evaporation and chelating agent were completed at 350 °C. Based on those results, the crystallization behavior of ***ink 12-0.1-60*** was studied using thermal XRD at temperatures above 350°C. Figure 4 shows the XRD patterns performed on powder synthesized by calcining the ink at 400°C and 500°C for 1h.

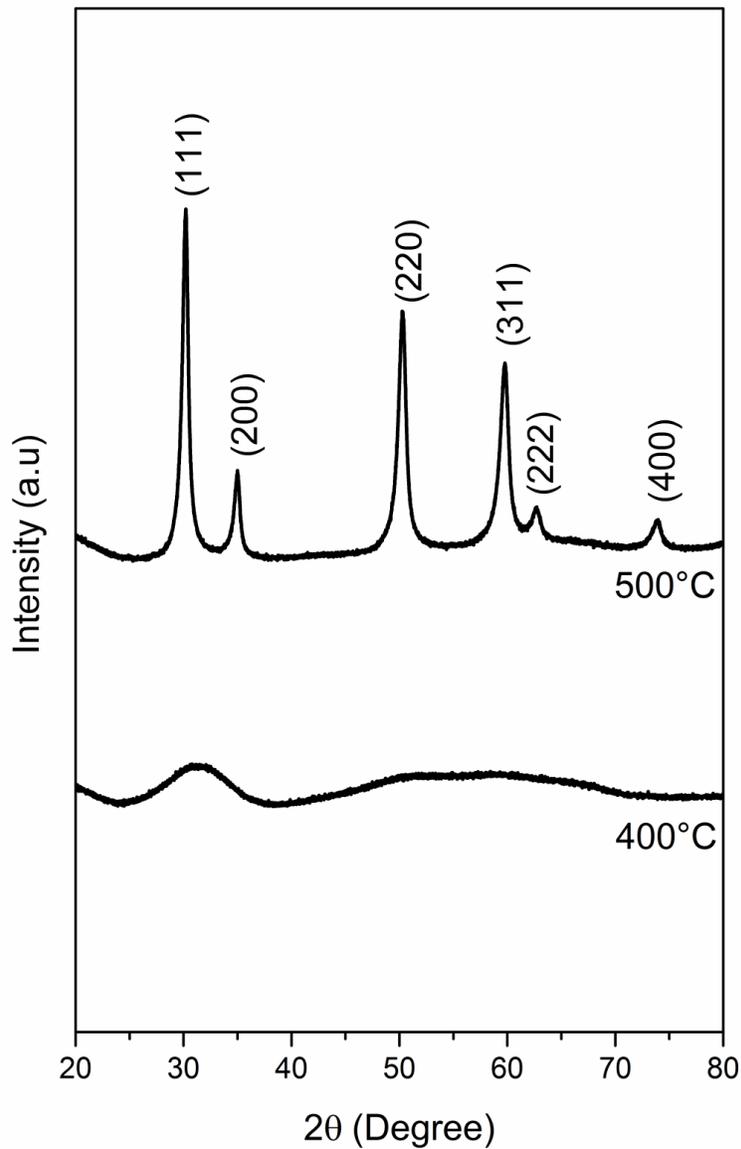

**Figure 4: XRD pattern of fresh ink 12-0.1-60 at different temperatures.**

Diffraction patterns show that no clear crystalline formation could be observed for the powder calcined at 400°C, while a well-defined pattern was observed at 500°C. The pattern observed at 500°C indicates that the formed material is indeed 8YSZ [48], with a crystallite size evaluated at 15 nm *via* Scherrer equation. In the light of such results, the calcination temperature of 500°C was chosen for the firing of the printing.

Figure 5 shows the microstructures of the film deposited with the thermal HP printer after calcination at 500°C.

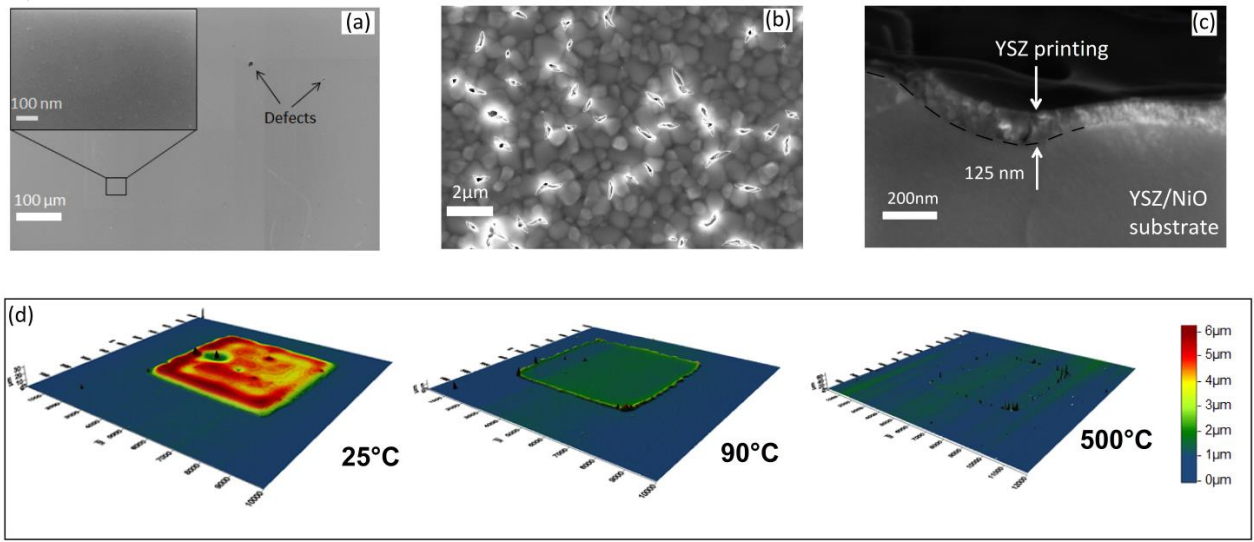

**Figure 5: SEM images of the printed fresh ink 12-0.1-60 on glass calcined at 500°C for 1h at low (a) and high (inset) magnification, printed and calcined on NiO/YSZ from surface (b) and cross section view (c), and 3D reconstruction of profilometry data at different temperatures (d).**

Particularly, Figure 5 (a) shows the surface of calcined *ink 12-0.1-60* on dense glass substrate at low magnification after calcination. The resulting film is homogeneous and continuous, with a few defects probably due to the inaccuracy of the used HP printer [15]. Microstructural features could only be observed at very high magnification (inset Figure 5 (a)) and exhibits a very dense microstructure with very small grains of ca. 15 nm. The crystallite size estimation by SEM pictures was consistent with the crystallite size obtained from XRD measurements. The film thickness was estimated by SEM of sample cross sections while the evolution of the thickness with the calcination was measured by using 3D profilometry. The printing thickness was found slightly dependent on the substrate porosity with average values below 150 nm for one-layer printing, also confirmed by laser profilometry. Figure 5 (b) shows the typical microstructure of the printed electrolyte layer on a porous NiO/YSZ SOFC anode. Some notable differences with the deposition on glass (Figure 5 (a)) are readily noticed. Due to the porosity of the NiO/YSZ surface, the resulting film after calcination shows a relatively high number of cracks, all located above pores. The presence of such cracks can possibly be attributed to the following two effects:

- First, the combination of large pores and very low film thickness, (measured on a cross section (Figure 5 (c)) at around 100 nm) might induce film cracks due to capillary forces.
- Secondly, the film shrinkage during calcination could also induce these cracks. Sol gel based solutions contain a high amount of solvent and volatile elements and it results in very high volume shrinkage during heat treatment [49]. Note however, that the cracks were not observed with the flat, pore-free soda-lime glass substrate (Figure 5(a)).

The shrinkage of this ink has been estimated using 3D profilometry on a 5×5 mm$^2$ print of fresh ***ink 12-0.1-60*** by measuring the print dimensions (length *x*, width *y*, and thickness *z*) at different stages of heat treatment. 3D images were then reconstructed and are shown on Figure 5(d). No shrinkage was measured in the lateral (i.e. *x*- and *y*-) directions, even after calcination at 500°C. However, a remarkable shrinkage of 90 vol% volume reduction after calcination was measured in the vertical (*z*-) direction. Such a high shrinkage, along with capillary pressure effect, will strongly participate to the collapsing of the film through the pores of the substrate. ~~Optimization of the substrate surface prior to~~

~~printing and modification of the printing process (multilayer deposition, intermediate drying step for example) could help to enhance the quality of thin film electrolyte.~~

*3.4. Electrical characterization*

For the electrical characterization, optimized **ink 12-0.1-60** was printed onto quartz and alumina substrates using the thermal inkjet printer. The ionic conductivity of the calcined thin film was measured by electrochemical impedance spectroscopy (EIS) in air between 500-750°C. The measurements were taken in the lateral configuration, *i.e.* in-plane electrodes deposited on the film surface (see drawings Figure 6a). This is an optimal electrode/electrolyte configuration for nanometric thin films due both to the maximization of electrolyte impedance with respect of the geometrical features (thickness is below 150 nm) and to the minimization of the electrodes contribution to the total impedance [39, 40]. Figure 6 shows a typical Nyquist and Bode plots of the film on quartz taken at 750 °C in air (Figure 6a) and the Arrhenius plot of the films on quartz and alumina calculated from the EIS data at different temperatures (Figure 6b). Recorded Nyquist plots on quartz exhibited a single semi-arc due to a dominant grain boundary contribution, i.e. a single R//C Voight element. This is modeled by a typical RQ circuit (solid line in Figure 6a) while the inset Figure 6a shows a Bode plot of the same data. The RQ circuit resistance was 9.00 MΩ and the summit frequency 972 Hz. The *n*-value for the constant phase element Q was 0.892. The calcined films on quartz resulted crack-free and homogeneous. The results are consistent with the average grains size of 15 nm measured by XRD and SEM. This electrical behavior is typical of nanocrystalline thin films [39,40] and the electrolyte ohmic resistance ($R_{eletrolyte}$ in Figure 6a) is calculated at the relaxation of the equivalent RQ circuit. Total conductivity and activation energy ($E_a$) were then calculated from the measured real impedance components ($R_{el}$) at different temperatures and normalized for the electrolyte/thin film geometry (Figure 6b).

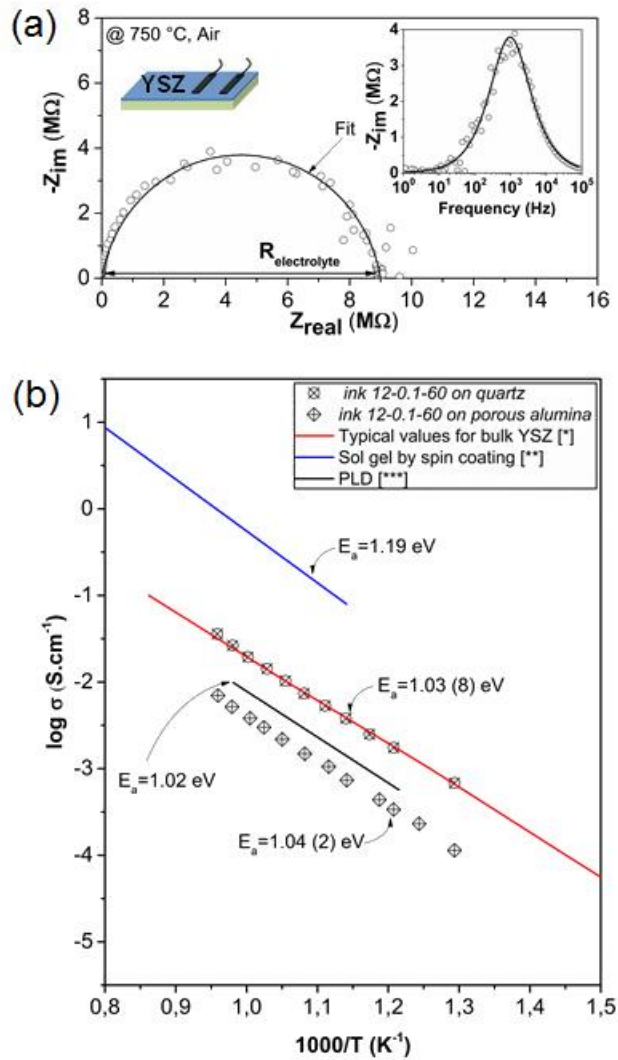

**Figure 6: (a) Typical Nyquist and Bode plots of the film on quartz taken at 750 °C in air and (b) Arrhenius plot of the films on quartz and alumina calculated from the EIS data at different temperatures along with typical conductivity values reported for YSZ thin film (red line, \*=[50]), and values reported for 8YSZ thin film deposited by sol gel spin coating (blue line, \*\*=[51]) and PLD (black line, \*\*\*=[52]).**

For comparison typical YSZ thin film conductivity measured by EIS [50] is shown in Figure6b (red line) as well as Arrhenius plots from previous studies of 8YSZ thin films deposited by sol-gel (spin coating) (blue line) [51], and PLD (black line) [52]. The calculated conductivity and activation energy (1.03 ±0.08 eV) of our inkjet printed thin film, are consistent with typical values reported for 8YSZ thin films. The conductivity at 750°C ($3.1 \cdot 10^{-2}$ S/cm) is as expected for bulk YSZ [50] and the inkjet printed thin film shows performances comparable with other films deposited by PLD [52]. Figure 6b also shows results collected on the alumina substrate. In that case the film was partially cracked, due to the roughness and asperities of the alumina substrates. Despite the lower conductance, the inkjet layer showed comparable conductivity with PLD films, indicating an overall reliability of the deposited layers by inkjet on rough substrates.

## Conclusion

Stable aqueous-based metal-organic solutions for inkjet printing were designed and demonstrated for printing of 8YSZ thin films and micrometric patterns. Ink printability and long term stability was optimized by the adjustment of the nucleophilic agent, metal oxide concentration and solvent composition. An optimized ink with $c = 0.1$ g/ml, [MDEA]/ [Zr(OPr)4] = 12 and water/ethanol = 60/40 vol% shows remarkable long term stability of 360 h, with a printability Z number around 1, and an exceptional jettability after 1200 h of storage. Nanocrystalline 8YSZ thin films of 100-150 nm were obtained after calcination at 500 °C on different substrates including Si, quartz, polycrystalline $Al_2O_3$ and tape cast NiO/YSZ SOFC anode substrate. The measured ionic conductivity agrees well with previously reported 8YSZ thin-film conductivity data. Formation of cracks in the printed films mainly depends on the quality of the substrate.

## Acknowledgement

The authors would like to thank Mads Gudik-Sørensen (DTU) and Victor Costa Bassetto (EPFL) for their technical assistance. This project has received funding from the Fuel Cells and Hydrogen 2 Joint Undertaking under grant agreement No 700266. This Joint Undertaking receives support from the European Union´s Horizon 2020 research and innovation program and Hydrogen Europe and N.ERGHY.

**Figure 1:** Flow curve and viscosity curves (inset) of the four developed inks measured 1 h after preparation.

**Figure 2:** Inks printability at different concentrations and aging time *versus* complexation ratio.

**Figure 3:** Droplet at different nozzle distances (frequency=1 kHz) of ink 12-0.1-60 as prepared (a), aged for 48h (b), and aged for 1200h (c), followed by droplets of fresh ink 12-0.1-60 (frequency = 10 kHz) (d) and optical pictures of an array of fresh ink 12-0.1-60 printed on NiO/YSZ (e) and alumina (f).

**Figure 4:** XRD pattern of fresh ink 12-0.1-60 calcined at different temperatures.

**Figure 5:** SEM images of the printed fresh ink 12-0.1-60 on glass calcined at 500°C for 1h at low (a) and high (inset) magnification, printed and calcined on NiO/YSZ from surface (b) and cross section view (c), and 3D reconstruction of profilometry data at different temperatures (d).

**Figure 6:** (a) Typical Nyquist and Bode plots of the film on quartz taken at 750 °C in air and (b) Arrhenius plot of the films on quartz and alumina calculated from the EIS data at different temperatures along with typical conductivity values reported for YSZ thin film (red line, *=50), and values reported for 8YSZ thin film deposited by sol gel spin coating (blue line, **=51) and PLD (black line, ***=52).

**Table 1: table viscosity and printability values**

|  | $t=1h$ | | | $t=48h$ | | | $t=192h$ | | | $t=360h$ | | |
|---|---|---|---|---|---|---|---|---|---|---|---|---|
|  | $\eta$ | $\sigma$ | $Z$ | $\eta$ | $\sigma$ | $Z$ | $\eta$ | $\sigma$ | $Z$ | $\eta$ | $\sigma$ | $Z$ |
|  | mPa·s | mN/m | - | mPa·s | mN/m | - | mPa·s | mN/m | - | mPa·s | mN/m | - |
| **Ink 9-0.1-100** | 9.4 | 33.4 | 2.83 | 11.5 | 33.6 | 2.32 | 37 | 33.5 | 0.72 | 148 | 33.3 | 0.18 |
| **Ink 10-0.1-100** | 10.6 | 33.1 | 2.50 | 13.1 | 32.9 | 2.02 | 18.3 | 33.4 | 1.45 | 25.9 | 33.2 | 1.02 |
| **Ink 12-0.1-100** | 15.2 | 31.4 | 1.70 | 16.7 | 31.1 | 1.54 | 16.7 | 31.7 | 1.55 | 18 | 31.2 | 1.43 |
| **Ink 15-0.1-100** | 28.1 | 31.3 | 0.92 | 29 | 31.5 | 0.89 | 26.5 | 31.2 | 0.97 | 28.9 | 31.6 | 0.90 |

**Table 2: viscosity/printability of ethanol diluted inks for ink 12-01-$X$**

| $X$ Water/Ethanol (vol%) | $t=1h$ | | | | $t=360h$ | | | |
|---|---|---|---|---|---|---|---|---|
|  | $\eta$ | $\sigma$ | $\rho$ | $Z$ | $\eta$ | $\sigma$ | $\rho$ | $Z$ |
|  | mPa·s | mN/m | g·cm$^{-3}$ | - | mPa·s | mN/m | g·cm$^{-3}$ | - |
| **100/0** | 15.2 | 31.4 | 1.06 | 1.70 | 18 | 31.7 | 1.06 | 1.43 |
| **80/20** | 15.1 | 30.3 | 1.03 | 1.65 | 17 | 30.1 | 1.03 | 1.47 |
| **60/40** | 14.9 | 29.5 | 1.01 | 1.64 | 16.7 | 29.1 | 1.01 | 1.46 |
| **40/60** | 12.8 | 27.3 | 0.99 | 1.82 | 14.8 | 27.8 | 0.99 | 1.57 |
| **20/80** | 12.5 | 26.1 | 0.96 | 1.79 | 11.3 | 26.4 | 0.96 | 1.98 |
| **0/100** | 7 | 24.5 | 0.94 | 3.07 | 8.01 | 24.8 | 0.94 | 2.68 |